\documentclass[reprint, longbibliography, aps, nofootinbib, showpacs, superscriptaddress, prb, preprintnumbers]{revtex4-2}
\pdfoutput=1
\usepackage[utf8]{inputenc}
\usepackage{mathtools}
\usepackage{hyperref,amsmath,amssymb,physics,graphicx}
\usepackage{caption}
\usepackage{subcaption}
\usepackage{orcidlink}

\usepackage{listings}
\usepackage{color}
\definecolor{dkgreen}{rgb}{0,0.6,0}
\definecolor{gray}{rgb}{0.5,0.5,0.5}
\definecolor{mauve}{rgb}{0.58,0,0.82}

\usepackage{xcolor}

\newtheorem{theorem}{Theorem}[section]

\newtheorem{conjecture}[theorem]{Conjecture}

\begin{document}
    
    \title{One-half reflected entropy is not a lower bound for
entanglement of purification}
    
    \author{Josiah Couch\orcidlink{0000-0002-7416-5858}}
    \affiliation{Beth Israel Deaconess Medical Center,
        Boston, MA, 02215, USA}
    \author{Phuc Nguyen\orcidlink{0000-0001-9993-8434}}
    \affiliation{Brandeis University, Waltham, MA 02453, USA}
    \author{Sarah Racz \orcidlink{0000-0002-4440-6163}}
    \affiliation{Theory Group, Weinberg Institute, Department of Physics, The University of Texas at Austin, Austin, TX 78712, USA}
    \author{Georgios Stratis \orcidlink{0000-0002-1346-8417}}
    \affiliation{Department of Electrical and Computer Engineering, Northeastern University, Boston MA, 02115, USA}
    \author{Yuxuan Zhang\orcidlink{0000-0001-5477-8924}}
    \thanks{quantum.zhang@utoronto.ca}
    \affiliation{Department of Physics, The University of Texas at Austin, Austin, TX 78712, USA}
    \affiliation{Department of Physics and Centre for Quantum Information and Quantum Control, University of Toronto,
60 Saint George St., Toronto, Ontario M5S 1A7, Canada}
    \affiliation{Vector Institute, MaRS Centre, Toronto, Ontario, M5G 1M1, Canada}
\begin{abstract}
    In recent work, Akers et al. proved that the entanglement of purification $E_p(A:B)$ is bounded below by half of the $q$-R\'enyi reflected entropy $S_R^{(q)}(A:B)$ for all $q\geq2$, showing that $E_p(A:B) = \frac{1}{2} S_R^{(q)}(A:B)$ for a class of random tensor network states. Naturally, the authors raise the question of whether a similar bound holds at $q = 1$. Our work answers that question in the negative by finding explicit counter-examples, which we arrive at through numerical optimization.
    Nevertheless, this result does not preclude the possibility that restricted sets of states, such as CFT states with semi-classical gravity duals, could obey the bound in question.
\end{abstract}
    
\maketitle


\section{Introduction}

One of the major thrusts of work towards understanding the AdS/CFT correspondence has been the program to map certain geometric properties of the AdS bulk with information-theoretic quantities on the CFT boundary. One of these geometric quantities of interest has been the entanglement wedge cross-section $EW$, which was first conjectured to be dual to the entanglement of purification \cite{Takayanagi:2017knl, Nguyen:2017yqw}, before being shown to be dual to ($\frac{1}{2}$ times) a newly defined quantity termed the reflected entropy \cite{Dutta:2019gen}. Given this history, one may be left wondering what the precise relationship between entanglement of purification and reflected entropy is\footnote{For additional work related to these conjectures, see \cite{Bao:2017nhh, Hirai:2018jwy, Bao:2018gck, Umemoto:2018jpc, Kudler-Flam:2018qjo, Tamaoka:2018ned, Bao:2018fso, Caputa:2018xuf, BabaeiVelni:2019pkw, Umemoto:2019jlz, Bao:2019zqc, Akers:2019gcv,velni2020evolution, Li:2020ceg, Sahraei:2021wqn, Bao:2021vyq, Vasli:2022kfu, Afrasiar:2022wzn, velni2023entanglement}}. One step towards answering this question was recently made by Akers et al. \cite{Akers:2023obn}, who found that entanglement of purification is bounded below by half the 2-R\'enyi version of reflected entropy for all quantum states. A question raised in that work is whether this can be strengthened to a bound by the ordinary (i.e., 1-R\'enyi) quantity. This work will present counter-examples showing that the stronger bound does not hold.

\begin{figure}
        \centering
        \includegraphics[width=.55\linewidth]{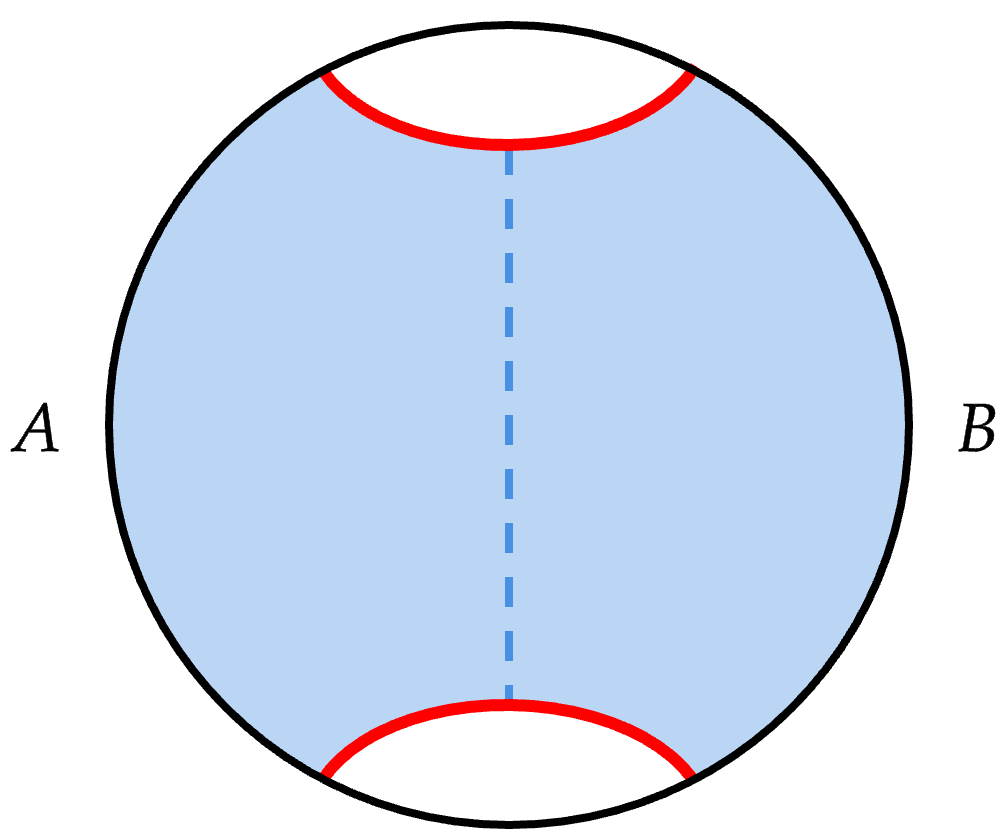}
    \caption{The entanglement wedge of subregions $A$ and $B$ is shown shaded in blue is bounded by $A$ and $B$ as well as the RT surface (shown in red). The entanglement wedge cross section, shown by a blue dashed line, is the minimal surface that separates A from B.}
    \label{fig:EWCS}
\end{figure}

In holographic systems, the Ryu-Takayanagi (RT) formula relates the entanglement entropy of states to a minimal surface in the bulk \cite{Ryu:2006bv} called the RT surface. The region enclosed by this minimal surface and the physical boundary is called the entanglement wedge. The minimal cut that bisects the entanglement wedge is what we refer to as the entanglement wedge cross-section, EW. An example illustrating EW geometrically is shown in Fig. \ref{fig:EWCS}.

The entanglement of purification $E_{p}(A:B)$ of a bipartite quantum system was defined in \cite{Terhal_2002} as follows: Let $\rho_{AB}$ be a bipartite density matrix. Consider all purification $\ket{\psi} \in \mathcal{H}_{AA'} \otimes \mathcal{H}_{BB'}$ of $\rho_{AB}$ on four parties: $E_{p}(A:B)$ is the smallest von Neumann entropy of $AA'$ among all such purifications. In other words:
\begin{equation}
    E_{P}(A:B)= \min_{\rho_{AB} = \mathrm{Tr}_{A'B'}(\ket{\psi} \bra{\psi})} S(AA')_{\ket{\psi}}\,,
\end{equation}
where $S(AA')$ denotes the von Neumann entropy of $\rho_{AA'}$.\footnote{Additional details about this quantity were worked out in \cite{Chen_2012, Bagchi_2015}} Reflected entropy, denoted by $S_R(A:B)$, on the other hand, is defined in \cite{Dutta:2019gen} as the entanglement entropy $S(AA')$ in a preferred 4-party purification of $\rho_{AB}$ in a doubled Hilbert space. The Gelfand–Naimark–Segal (GNS) construction, which provides a way to represent states on a $C^*$ algebras as states on a Hilbert space, defines states that are examples of such purifications \cite{Dutta:2019gen}. Unlike $E_p(A:B)$, for a given $\rho_{AB}$, $S_R(A:B)$ is defined through a canonical purification rather than a minimization procedure, making $S_R(A:B)$ computationally much simpler. The reflected entropy can be extended to a notion of $q$-R\'enyi reflected entropy by replacing the entanglement entropy $S(AA')$ by $q$-R\'enyi reflected entropy $S^q(AA')$. In other words, the reflected $q$-entropy is defined as:
\begin{eqnarray}
    S_R^{(q)}(A:B) = S^{(q)}({\rm Tr }_{BB'}\ketbra{\sqrt{\rho_{AB}}}{\sqrt{\rho_{AB}}})\,,
\end{eqnarray}
where $S^{(q)}$ is the $q$-R\'enyi entropy. Noticeably, Akers et al.~\cite{Akers:2023obn} proved that that for any bipartite mixed quantum state $\rho_{AB}$ and any integer $q\geq 2$,
\begin{equation}\label{eq:akers_bound}
    E_P(A:B) \geq \frac{1}{2} S^{(q)}_R(A:B)\,,
\end{equation}
where $S^{(q)}_R(A:B)$ is the reflected $q$-R\'enyi entropy of $\rho_{AB}$. A natural question raised in the discussion of that paper is whether this bound can be strengthened to include the $q=1$ case, as proving that would lead to a key advancement in proving the desired $E_p = EW$ relationship. In their work, the authors made the following conjecture:
\begin{conjecture} [Akers et al.] \label{conj:lb}
For all bipartite quantum states, 
\begin{align}\label{eq:strengthened_bound}
    E_P(A:B) \geq \frac{1}{2}S_R(A:B)
\end{align}

\end{conjecture}
However, this seems too good to be true for generic quantum states. First of all, recently, Hayden et al.~\cite{Hayden:2023yij} observed that the reflected entropy is not a good correlation measure for $0<q<2$, while the entanglement of purification is, due to the former's lack of monotonicity. Moreover, intuitively, the choice of purification in reflected entropy seems somewhat arbitrary: any unitary operation acting on the auxiliary space of the purified state, $\mathcal{H_{A'}}\otimes\mathcal{H_{B'}}$, creates a valid purification and would thus generally give rise to different entropy values. In this paper, we construct explicit counter-examples using numerical solvers, falsifying the conjecture.
\begin{theorem}\label{the:lb}
    There exists a bipartite Hilbert space $\mathcal{H_{A}}\otimes\mathcal{H_{B}}$ and a density operator $\rho_{AB}$ on that space such that  $$E_P(A:B) < \frac{1}{2}S_R(A:B).$$
\end{theorem}

We then study the properties of the states violating this inequality, as well as explore inequalities like Eq.~\ref{eq:akers_bound} for non-integer R\'enyi indices. We also discuss the differences in multi-party entanglement between holographic or tensor network states and generic quantum states.

In addition, some of our counter-examples can be shown to violate a similar bound for some non-integer $1<q<2$, though much closer to 1, as shown in Fig~\ref{fig:gap}. Combined with the observation made by Hayden et al.~\cite{Hayden:2023yij}, this leads us to make the following conjecture:
\begin{conjecture}\label{conj:vio}
    For any $q \in[0, 2)$, there exists a density operator $\rho_{AB}$ on $\mathcal{H_{A}}\otimes\mathcal{H_{B}}$ such that $$E_P(A:B) < \frac{1}{2}S_R^{(q)}(A:B).$$
\end{conjecture}
In the next section, we explain our numerical algorithm and results in detail.

\section{Numerical counter-examples}\label{sec: numerics}

We numerically search for counter-examples to inequality \ref{eq:strengthened_bound}. As the computation of $E_P(A:B)$ can be demanding, we design the following trick: by definition, the entanglement of purification minimizes the entropy of $AA'$ over all possible purifications and partitions; therefore, $any$ fixed purification and partition should be an upper bound of this quantity. It follows that any four-partite pure quantum state $\psi_{ABA'B'}$ satisfies: 
\begin{equation}
    S(AA') \geq E_P(A:B) \,.
\end{equation}
As such, inequality \ref{eq:strengthened_bound} implies the bound 
\begin{equation}
    S(AA') \geq \frac{1}{2} S_R(A:B) \,,
\end{equation}
for any such four-partite pure state. For numerical simplicity, it is this bound for which we find a counter-example. 

Our numerical code is written in Python using PyTorch~\cite{paszke2019pytorch} and the tensor network package~\cite{roberts2019tensornetwork}. The basic idea of the code is as follows: we define an initial 4-partite state $\ket{\chi_{ABA'B'}}$ to be the equal superposition state overall computational bases,\footnote{This choice is only for numerical stability} i.e.
\begin{gather}
    \ket{\chi_{ABA'B'}} \propto 
    \sum_{i,j,k,l=1}^{d_{A}, d_{B}, d_{A'}, d_{B'}}  \ket{i}_{A} \otimes \ket{j}_{B} \otimes \ket{k}_{A'} \otimes \ket{l}_{B'} 
\end{gather}
where $d_{A}, d_{B}, d_{A'},$ and $d_{B'}$ are the dimensions of the $A, B, A',$ and $B'$ systems respectively, and we then construct a unitary $U$ from an upper-triangular complex matrix $M$\footnote{We choose to represent our unitary matrices this way for computational convenience. It is the entries of the matrix $M$ that will serve as the parameters we will optimize over below.} using
\begin{equation}
    U = e^{M - M^\dagger} \,.
\end{equation}
We can then compute $S(AA')$ and $S_R(A:B)$ for 
\begin{equation}
    \ket{\psi_{ABA'B'}} = U \ket{\chi_{ABA'B'}}\,,
\end{equation}
and minimize 
\begin{equation}
    S(AA') - \frac{1}{2} S_R(A:B) \,,
    \label{eq:delta_fn}
\end{equation}
in the state $\ket{\psi_{ABA'B'}}$ with respect to the entries of the upper triangular matrix $M$ using the built-in PyTorch implementation of the ADAM~\cite{kingma2017} optimizer. 

\begin{figure}
    \centering
    \begin{subfigure}[a]{0.45\textwidth}
        \centering
        \includegraphics[width=1\linewidth]{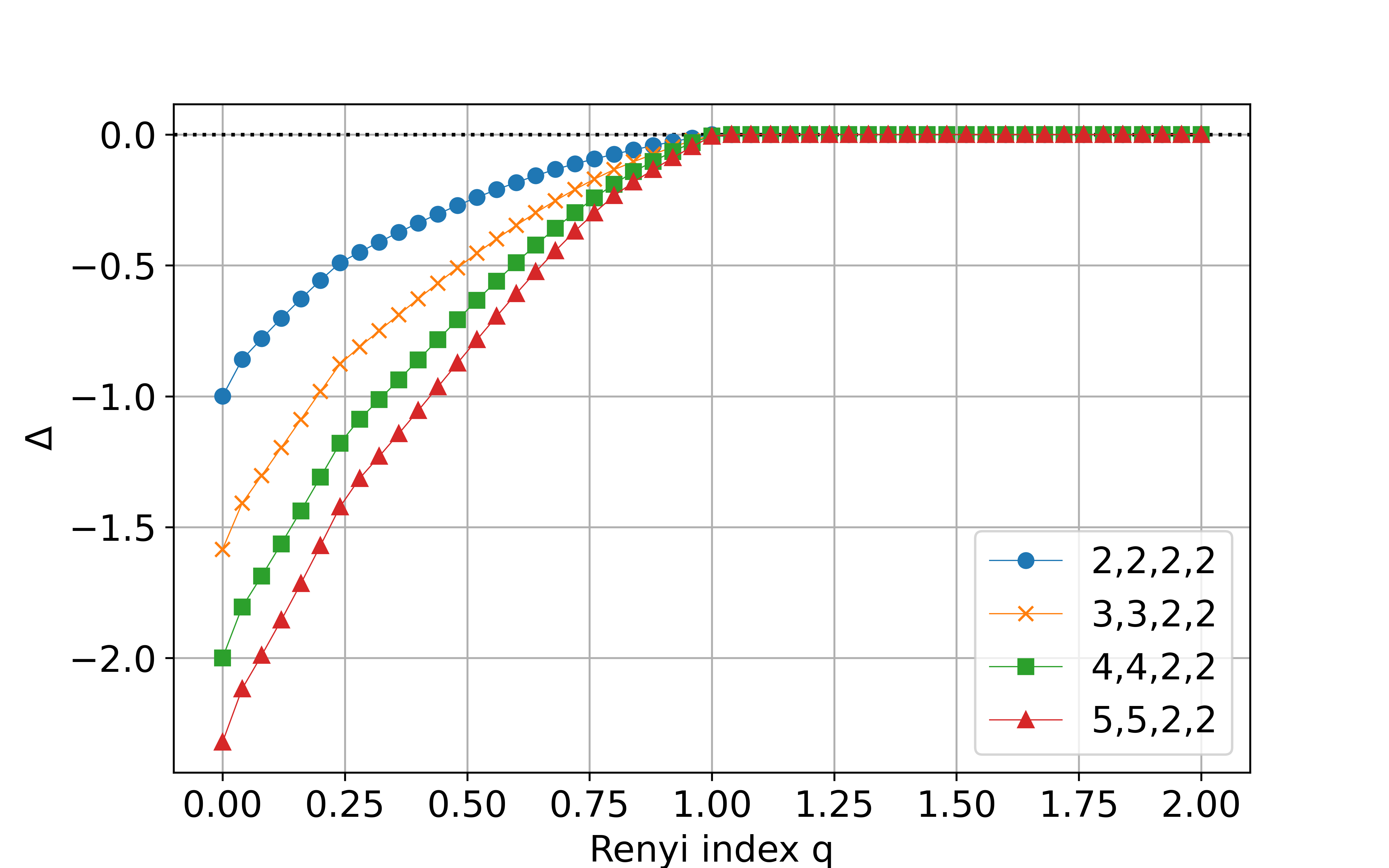}
    \end{subfigure}
    \begin{subfigure}[b]{0.45\textwidth}
        \centering
        \includegraphics[width=1\linewidth]{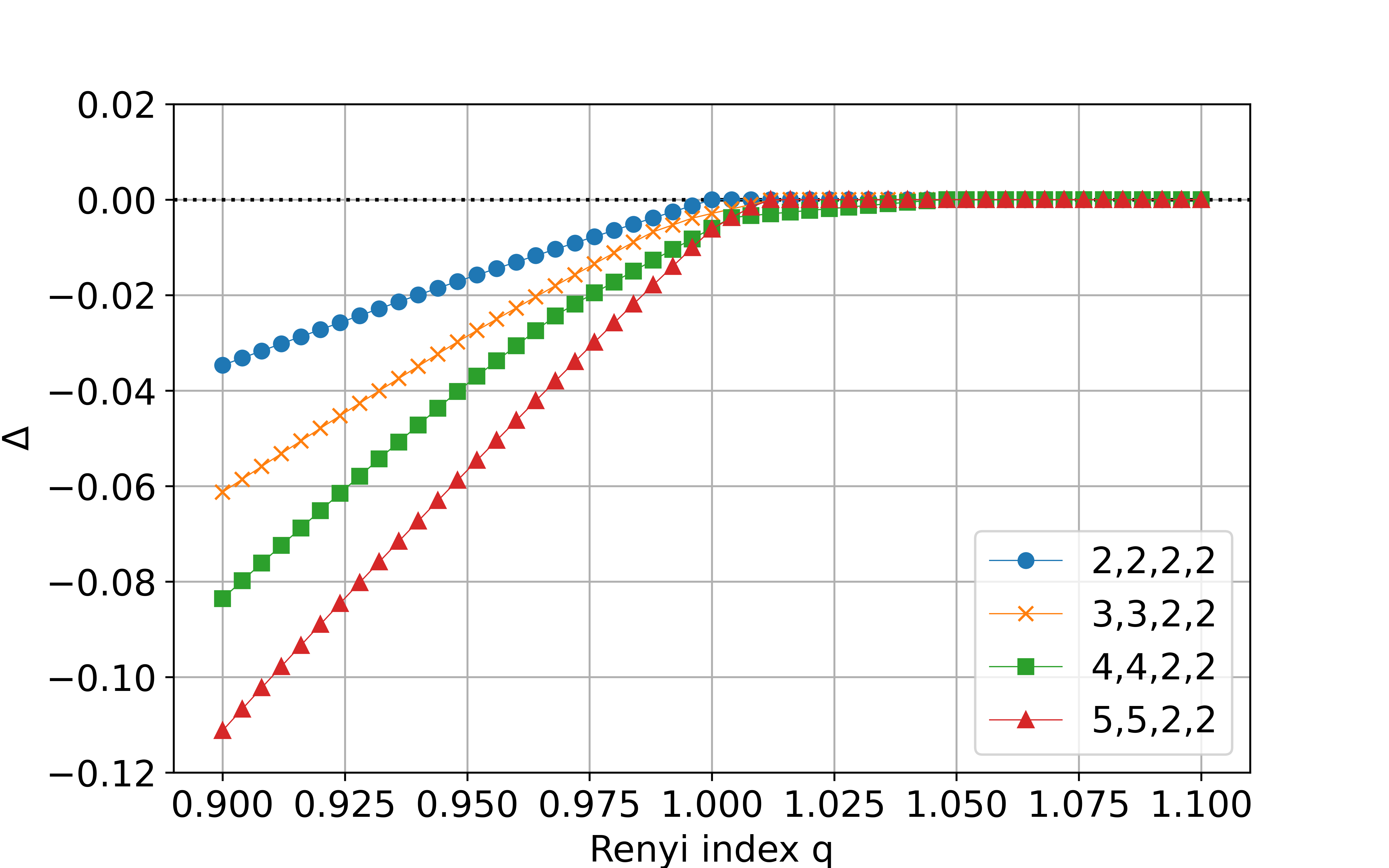}
    \end{subfigure}
    \caption{In order to study the lower bound violation at different Renyi indices, $q$, we first run a batch of optimizations at several $q$'s (0.1, 0.5, 0.9, 0.99, 1, and 1.02). Then, to generate a smooth curve, we take all the resultant states and calculate their respective $\Delta$ at various values of $q$'s. In the end, for each system size and each $q$, we plot the smallest $\Delta$ in the set. The bottom panel is a zoom-in of the top panel at $q\in[0.9,1.1]$. We denote with circle, `x', square, and triangle  markers the [2, 2, 2, 2], [3, 3, 2, 2], [4, 4, 2, 2], and [5, 5, 2, 2] systems respectively.}
    \label{fig:gap}
\end{figure}

\begin{figure}
    \centering
    \begin{subfigure}[a]{0.48\textwidth}
        \centering
        \includegraphics[width=1\linewidth]{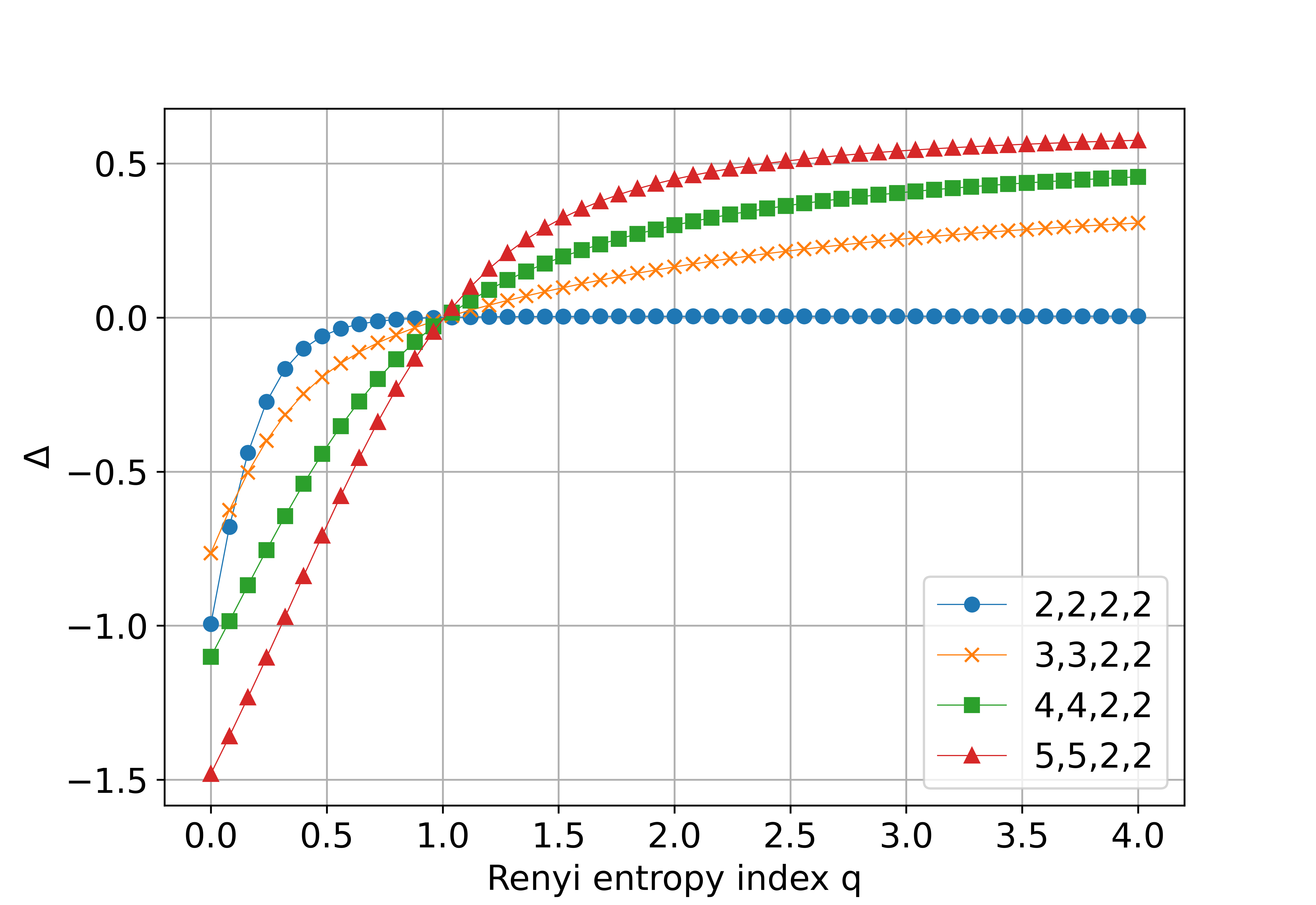}
        \label{fig:gap_v_q_single_state1}
    \end{subfigure}
    \begin{subfigure}[b]{0.48\textwidth}
        \centering
        \includegraphics[width=1\linewidth]{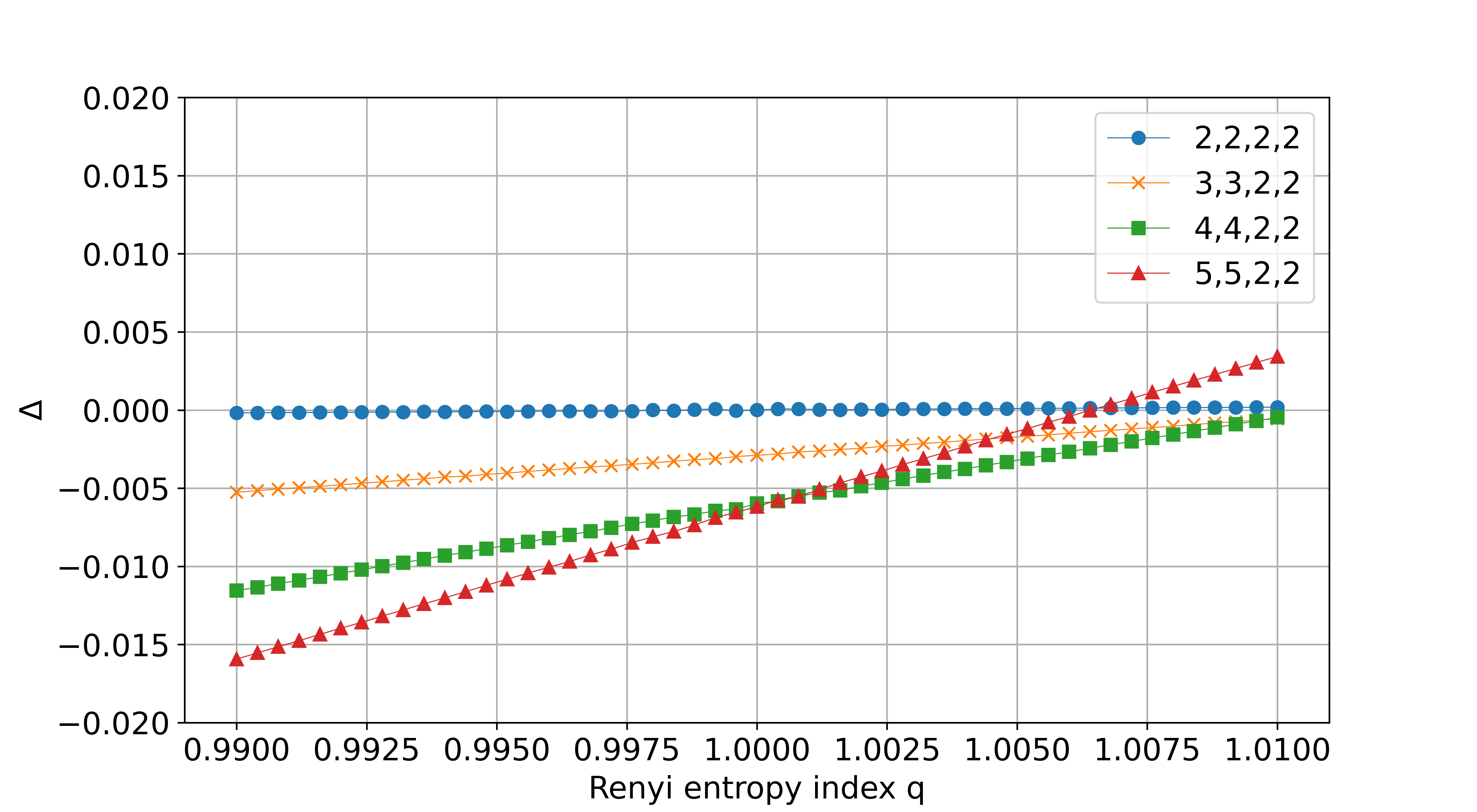}
        \label{fig:gap_v_q_single_state2}
    \end{subfigure}
    \caption{We look at $\Delta$ versus q for a single state at each dimension. The bottom panel is once again zoomed in on the top panel. Notice that if we were to plot the minimum over all possible states, then fixing $A'B'$, the $\Delta$ function from a smaller $AB$ should always be the upper bound of that of a larger one. However, this is not the case if we look at a single state. Notice that we have numerical noise in a neighborhood around $q=1$, which we believe is due to the fact that the Renyi entropy at $q=1$ is defined only through a limit. The curves follow the same color code as Fig.~\ref{fig:gap} for consistency. 
    }
\end{figure}

We define the entanglement gap function, denoted as $\Delta (q)$, between $S(AA')$ and $\frac{1}{2} S_R^{(q)}(A:B)$:
\begin{align}\label{eq:gap}
    {\rm \Delta} \coloneqq S(AA') - \frac{1}{2} S_R^{(q)}(A:B)\,.
\end{align}
Conj.~\ref{conj:lb} implies that for all quantum states, the $\Delta$ function ~\ref{eq:gap} should remain non-negative at $q = 1$, which we find is not the case. Our code produces counter-examples to inequality \ref{eq:strengthened_bound} of a number of different on-site dimensions (i.e. different choices of $d_{A}, d_{B}, d_{A'},$ and $d_{B'}$). We have included two examples in Append.~\ref{app:state} with on-site dimensions $d_{A}=d_{B}=3$, $d_{A'}=d_{B'}=2$ (which we denote [3,3,2,2]) for the first state and $d_{A}=d_{B}=4$, $d_{A'}=d_{B'}=2$ (denoted [4,4,2,2]) for the second. Likewise, we also generate counter-examples to conjecture~\ref{conj:lb} for on-site dimensions $[4,4,2,2]$ and $[5,5,2,2]$ while examining a few non-integer Renyi indices. Examining the $\Delta$ function in Fig.~\ref{fig:gap}, two trends emerge: First, $\Delta$ stays negative and grows monotonically between $(0,1]$ until it closes to 0 at $q\sim 1.04$; secondly, without increasing the size of $A'B'$, increasing the on-site dimensions on $AB$ tends to increase $\Delta$. The former is expected from the property of entropy and the latter could be intuitively understood as that the canonical purification always doubles the size of the Hilbert space. As such, it could generate more entanglement $S(AA')$ for a larger $AB$ size, whereas, by our design, two qubits always suffice to purify the state and the reduced density matrix of $AB$. Our work agrees with a recent observation \cite{Hayden:2023yij} that reflected entropy is ``not a correlation measure'' in that it does not decrease monotonically under partial trace for $0<q<2$. 

\section{Implications and discussions}\label{sec: discussion}

In this work, we have numerically constructed counter-examples when one-half reflected entropy fails to be a lower bound for entanglement of purification at Renyi index $q\in(0,1.04]$, refuting Conj~\ref{conj:lb} proposed in~\cite{Akers:2023obn}. Notably, our method avoids the extensive computational resource requirement in calculating the entanglement of purification by upper bounding it with the bipartite entanglement entropy of a certain purification. Looking at the original motivation of Conj~\ref{conj:lb}, it is not obvious that the inequality will fail if we constrain to purifications with holographic duals as without that restriction the original $E_p = EW$ conjecture is ill-defined.

Therefore, we want to examine whether the bound-violating quantum states we found are holographic. Holographic states obey a special information-theoretic property: the tripartite mutual information (TMI) among arbitrary disjoint spatial regions should always remain negative~\cite{Hayden:2011ag}. The TMI for a tripartite system is defined as the following:  
\begin{align}
    I_3(A:B:C) \coloneqq I_2(A:B) + I_2(B:C) - I_2(B:AC) \,,
\end{align}
where $I_2(A:B) \coloneqq  S(A)+ S(B) - S(AB)$ is the bipartite mutual information which measures two-interval correlations. For each four-party state, there are four possible TMIs one could define. We computed the maximum of the possible TMIs, denoted as ${\rm Max}(I_3)$, for the bound-violating states with $ \Delta < 0$ found in Sec.\ref{sec: numerics} at $q = 1$ and for all states the maximum TMI was positive. Positive $I_3$ implies a ``classical'' correlation behavior between the multiplets \cite{caceffo2023negative} existing in each of the bound-violating states.

For all states, we observe positive ${\rm Max}(I_3)$, which implies that these states lie outside the holographic entropy cone. On the other hand, the counter-examples in~\cite{Hayden:2023yij} with non-monotonically decreasing reflected entropy correspond to classical probability distributions. In some sense, this means the states we found share a property reminiscent of these counter-examples in~\cite{Hayden:2023yij}, although, for the latter, we could not observe a negative $\Delta$ function at $q=1$.

To examine explicitly whether the bound violation exists for quantum states with holographic duals, we modify the objective $\Delta$ function in Eq.~\ref{eq:gap} by adding a penalty term, $\Gamma$:
$$\Gamma \coloneqq {\rm Max}(I_3) \ \ \ {\rm if}\ \ \ {\rm Max}(I_3) > 0$$ Namely, we penalize any positive TMI. Interestingly, our optimizer did not yield a negative $\Delta$ for these penalized optimizations at any of the previously examined system sizes.

We also construct holographic quantum states from a type of tensor-network construction called the multi-scale entanglement renormalization ansatz (MERA)~\cite{vidal2007entanglement}. MERA is a construction that generates a $(D+1)$-dimensional holographic geometry and has been widely studied and implemented on quantum computers~\cite{vidal2008class,evenbly2014scaling,anand2022holographic}. Despite using a similar numerical technique to the one in Sec.~\ref{sec: numerics}, our searches on 8- and 16-qubit MERAs were unable to yield a negative entanglement $\Delta$ at $q=1$.

From these two pieces of evidence we conclude that even though our work refutes Conj~\ref{conj:lb} for generic quantum states, the relationship between reflected entropy and entanglement of purification could remain valid for holographic states. 
That is to say, the modified version of Conj~\ref{conj:lb}:
\begin{conjecture}\label{conj:lb_holo}
For all bipartite quantum states with holographic duals, 
\begin{align}
    E_P(A:B) \geq \frac{1}{2}S_R(A:B) \,
\end{align}
\end{conjecture}
could hold.

Moreover, it would be desirable to build an efficient numerical optimizer to search for the entanglement of purification, with which one could provide clues for future theoretical investigation.

For data and our full Python code, please visit \url{https://github.com/IosiaLectus/EofP_minus_half_RE_opt}.
\section*{Acknowledgements}

We would like to thank Chris Akers, Alex May, Pratik Rath, Brian Swingle, and Jonathan Sorce for their valuable comments and suggestions regarding this project. PN is supported by the Air Force Office of Scientific Research under award number FA9550-19-1-0360, by the U.S. Department of Energy through DE-SC0009986, and by the Simons Foundation via the
It From Qubit collaboration. SR is supported by the National Science Foundation under NSF Award No. 2210562. YZ is supported by the U.S. Department of Energy DOE DE-SC0022102 and by the Centre for Quantum
Information and Quantum Control (CQIQC) at the University
of Toronto. YZ thanks Timothy Hsieh and the Perimeter Institute for Theoretical Physics for hospitality, where part of the work is finished.
\newpage
\bibliographystyle{apsrev4-2}
\bibliography{main}
\newpage
\appendix
\onecolumngrid
\section{Examples of bound-violating quantum states}\label{app:state}
In the appendix we print out two pure states whose reduced density matrices on the first two qudits violate Conj.~\ref{conj:lb}. The first example has on-site Hilbert size [3,3,2,2,]:

\begin{eqnarray}
\psi_{ABA'B'} &=& \nonumber \\
  && (0.13 + 0.18 i )\ket{0, 0, 0, 0} + (0.073 + 0.259 i )\ket{0, 0, 0, 1} \nonumber \\
 &+& (0.091 + 0.229 i )\ket{0, 0, 1, 0} + (0.047 + 0.273 i )\ket{0, 0, 1, 1}   \nonumber \\
 &+& (-0.122 + 0.198 i )\ket{0, 1, 0, 0} + (-0.006 + 0.045 i )\ket{0, 1, 0, 1} \nonumber \\
 &+& (-0.049 + 0.175 i )\ket{0, 1, 1, 0} + (0.029 - 0.008 i )\ket{0, 1, 1, 1} \nonumber \\
 &+& (0.069 - 0.001 i )\ket{0, 2, 0, 0} + (-0.163 + 0.061 i )\ket{0, 2, 0, 1} \nonumber \\
 &+& (0.128 - 0.011 i )\ket{0, 2, 1, 0} + (-0.062 + 0.05 i )\ket{0, 2, 1, 1} \nonumber \\
 &+& (0.072 + 0.037 i )\ket{1, 0, 0, 0} + (0.061 + 0.119 i )\ket{1, 0, 0, 1} \nonumber \\
 &+& (-0.101 - 0.05 i )\ket{1, 0, 1, 0} + (-0.158 - 0.067 i )\ket{1, 0, 1, 1} \nonumber \\
 &+& (-0.117 + 0.181 i )\ket{1, 1, 0, 0} + (0.01 + 0.106 i )\ket{1, 1, 0, 1} \nonumber \\
 &+& (-0.175 - 0.169 i )\ket{1, 1, 1, 0} + (-0.101 + 0.033 i )\ket{1, 1, 1, 1} \nonumber \\
 &+& (0.003 + 0.047 i )\ket{1, 2, 0, 0} + (-0.159 + 0.103 i )\ket{1, 2, 0, 1} \nonumber \\
 &+& (-0.101 + 0.022 i )\ket{1, 2, 1, 0} + (-0.097 - 0.164 i )\ket{1, 2, 1, 1} \nonumber \\
 &+& (0.041 - 0.109 i )\ket{2, 0, 0, 0} + (-0.028 - 0.067 i )\ket{2, 0, 0, 1} \nonumber \\
 &+& (0.041 - 0.035 i )\ket{2, 0, 1, 0} + (0.001 + 0.03 i )\ket{2, 0, 1, 1} \nonumber \\
 &+& (-0.204 - 0.155 i )\ket{2, 1, 0, 0} + (-0.152 + 0.073 i )\ket{2, 1, 0, 1} \nonumber \\
 &+& (-0.172 + 0.022 i )\ket{2, 1, 1, 0} + (-0.069 + 0.103 i )\ket{2, 1, 1, 1} \nonumber \\
 &+& (-0.178 - 0.02 i )\ket{2, 2, 0, 0} + (-0.205 - 0.162 i )\ket{2, 2, 0, 1} \nonumber \\
 &+& (-0.093 + 0.034 i )\ket{2, 2, 1, 0} + (-0.202 - 0.014 i )\ket{2, 2, 1, 1} 
\end{eqnarray}

In this state, we compute
\begin{eqnarray}
    &S(AA')                        &\approx  0.81941\\
    &S_R(A:B)                      &\approx  1.64454\\
    &S(AA') - \frac{1}{2} S_R(A:B) &\approx -0.00286
\end{eqnarray}
And another example with on-site Hilbert size [4,4,2,2,]
\begin{eqnarray}
    \psi_{ABA'B'} &=& \nonumber \\
  && (0.078 + 0.055 i )\ket{0, 0, 0, 0, 0, 0} + (0.058 + 0.069 i )\ket{0, 0, 0, 0, 0, 1} \nonumber \\
 &+& (0.011 + 0.05 i )\ket{0, 0, 0, 0, 1, 0} + (-0.063 + 0.103 i )\ket{0, 0, 0, 0, 1, 1} \nonumber \\
 &+& (0.079 + 0.035 i )\ket{0, 0, 0, 1, 0, 0} + (0.004 + 0.06 i )\ket{0, 0, 0, 1, 0, 1} \nonumber \\
 &+& (-0.037 + 0.056 i )\ket{0, 0, 0, 1, 1, 0} + (-0.093 + 0.014 i )\ket{0, 0, 0, 1, 1, 1} \nonumber \\
 &+& (-0.011 - 0.029 i )\ket{0, 0, 1, 0, 0, 0} + (-0.039 - 0.077 i )\ket{0, 0, 1, 0, 0, 1} \nonumber \\
 &+& (-0.096 - 0.006 i )\ket{0, 0, 1, 0, 1, 0} + (-0.033 - 0.21 i )\ket{0, 0, 1, 0, 1, 1} \nonumber \\
 &+& (-0.034 - 0.083 i )\ket{0, 0, 1, 1, 0, 0} + (-0.097 - 0.07 i )\ket{0, 0, 1, 1, 0, 1} \nonumber \\
 &+& (-0.027 - 0.212 i )\ket{0, 0, 1, 1, 1, 0} + (-0.149 - 0.107 i )\ket{0, 0, 1, 1, 1, 1} \nonumber \\
 &+& (0.002 - 0.103 i )\ket{0, 1, 0, 0, 0, 0} + (-0.058 - 0.032 i )\ket{0, 1, 0, 0, 0, 1} \nonumber \\
 &+& (0.027 - 0.081 i )\ket{0, 1, 0, 0, 1, 0} + (0.013 - 0.071 i )\ket{0, 1, 0, 0, 1, 1} \nonumber \\
 &+& (-0.063 - 0.056 i )\ket{0, 1, 0, 1, 0, 0} + (-0.083 - 0.06 i )\ket{0, 1, 0, 1, 0, 1} \nonumber \\
 &+& (0.018 - 0.051 i )\ket{0, 1, 0, 1, 1, 0} + (0.004 - 0.104 i )\ket{0, 1, 0, 1, 1, 1} \nonumber \\
 &+& (-0.219 - 0.005 i )\ket{0, 1, 1, 0, 0, 0} + (-0.046 - 0.224 i )\ket{0, 1, 1, 0, 0, 1} \nonumber \\
 &+& (-0.05 - 0.134 i )\ket{0, 1, 1, 0, 1, 0} + (0.148 - 0.05 i )\ket{0, 1, 1, 0, 1, 1} \nonumber \\
 &+& (-0.066 - 0.218 i )\ket{0, 1, 1, 1, 0, 0} + (-0.226 - 0.015 i )\ket{0, 1, 1, 1, 0, 1} \nonumber \\
 &+& (0.144 - 0.068 i )\ket{0, 1, 1, 1, 1, 0} + (-0.02 - 0.113 i )\ket{0, 1, 1, 1, 1, 1} \nonumber \\
 &+& (0.053 - 0.08 i )\ket{1, 0, 0, 0, 0, 0} + (0.011 - 0.11 i )\ket{1, 0, 0, 0, 0, 1} \nonumber \\
 &+& (-0.085 - 0.127 i )\ket{1, 0, 0, 0, 1, 0} + (-0.081 - 0.083 i )\ket{1, 0, 0, 0, 1, 1} \nonumber \\
 &+& (0.018 - 0.082 i )\ket{1, 0, 0, 1, 0, 0} + (0.086 - 0.096 i )\ket{1, 0, 0, 1, 0, 1} \nonumber \\
 &+& (-0.09 - 0.104 i )\ket{1, 0, 0, 1, 1, 0} + (-0.1 - 0.154 i )\ket{1, 0, 0, 1, 1, 1} \nonumber \\
 &+& (0.011 - 0.122 i )\ket{1, 0, 1, 0, 0, 0} + (0.166 + 0.039 i )\ket{1, 0, 1, 0, 0, 1} \nonumber \\
 &+& (-0.177 - 0.019 i )\ket{1, 0, 1, 0, 1, 0} + (0.059 - 0.126 i )\ket{1, 0, 1, 0, 1, 1} \nonumber \\
 &+& (0.145 + 0.019 i )\ket{1, 0, 1, 1, 0, 0} + (0.042 - 0.08 i )\ket{1, 0, 1, 1, 0, 1} \nonumber \\
 &+& (0.022 - 0.138 i )\ket{1, 0, 1, 1, 1, 0} + (-0.111 - 0.012 i )\ket{1, 0, 1, 1, 1, 1} \nonumber \\
 &+& (0.02 - 0.098 i )\ket{1, 1, 0, 0, 0, 0} + (0.026 - 0.034 i )\ket{1, 1, 0, 0, 0, 1} \nonumber \\
 &+& (0.043 - 0.033 i )\ket{1, 1, 0, 0, 1, 0} + (0.0 - 0.083 i )\ket{1, 1, 0, 0, 1, 1} \nonumber \\
 &+& (0.108 - 0.058 i )\ket{1, 1, 0, 1, 0, 0} + (0.095 - 0.098 i )\ket{1, 1, 0, 1, 0, 1} \nonumber \\
 &+& (0.084 - 0.082 i )\ket{1, 1, 0, 1, 1, 0} + (0.103 - 0.01 i )\ket{1, 1, 0, 1, 1, 1} \nonumber \\
 &+& (-0.074 - 0.133 i )\ket{1, 1, 1, 0, 0, 0} + (0.06 - 0.041 i )\ket{1, 1, 1, 0, 0, 1} \nonumber \\
 &+& (-0.032 - 0.082 i )\ket{1, 1, 1, 0, 1, 0} + (0.038 + 0.0 i )\ket{1, 1, 1, 0, 1, 1} \nonumber \\
 &+& (0.059 - 0.059 i )\ket{1, 1, 1, 1, 0, 0} + (-0.084 - 0.07 i )\ket{1, 1, 1, 1, 0, 1} \nonumber \\
 &+& (0.04 - 0.003 i )\ket{1, 1, 1, 1, 1, 0} + (-0.043 - 0.025 i )\ket{1, 1, 1, 1, 1, 1}
\end{eqnarray}
In this state, we compute
\begin{eqnarray}
    &S(AA')                        &\approx  0.89796\\
    &S_R(A:B)                      &\approx  1.80783\\
    &S(AA') - \frac{1}{2} S_R(A:B) &\approx -0.00596
\end{eqnarray}

\end{document}